# Controlled synthesis of the antiperovskite oxide superconductor $Sr_{3-x}SnO$

J N Hausmann*[1,2], M Oudah*[1], A Ikeda[1], S Yonezawa[1] and Y Maeno[1]

* Equal contribution

[1] Department of Physics, Graduate School of Science, Kyoto University, Kyoto 606-8502, Japan.

[2] Department of Chemistry, Faculty of Mathematics and Natural Sciences, Humboldt-Universität zu Berlin, Brook-Taylor-Strasse 2, Berlin 12489, Germany.

Corresponding author: Y Maeno (email: maeno@scphys.kyoto-u.ac.jp).

**Abstract.** A large variety of perovskite oxide superconductors are known, including some of the most prominent high-temperature and unconventional superconductors. However, superconductivity among the oxidation state inverted material class, the antiperovskite oxides, was reported just recently for the first time. In this superconductor, $Sr_{3-x}SnO$, the unconventional ionic state $Sn^{4-}$ is realized and possible unconventional superconductivity due to a band inversion has been discussed. Here, we discuss an improved facile synthesis method, making it possible to control the strontium deficiency in $Sr_{3-x}SnO$. Additionally, a synthesis method above the melting point of $Sr_3SnO$ is presented. We show temperature dependence of magnetization and electrical resistivity for superconducting strontium deficient $Sr_{3-x}SnO$ ($T_c \sim 5$ K) and for $Sr_3SnO$ without a superconducting transition down to 0.15 K. Further, we reveal a significant effect of strontium raw material purity on the superconductivity and achieve 40% increased superconducting volume fraction (~100%) compared to the highest value reported so far. More detailed characterisation utilising powder X-ray diffraction and energy-dispersive X-ray spectroscopy show that a minor cubic phase, previously suggested to be a $Sr_{3-x}SnO$, is SrO. The improved characterization and controlled synthesis reported herein enable detailed investigations on the superconducting nature and its dependency on the strontium deficiency in $Sr_{3-x}SnO$.

## 1. Introduction

Oxides with perovskite related structures have been playing a major role in the study of superconductivity in the last three decades. Many of the milestones in high temperature superconductivity such as La$_{2-x}$Ba$_x$CuO$_4$ [1,2], YBa$_2$Cu$_3$O$_{7-x}$ [3] and Bi$_2$Sr$_2$CaCu$_2$O$_{8+\delta}$ [4], as well as the unconventional superconductor Sr$_2$RuO$_4$ [5] with evidence for spin-triplet superconductivity [6,7] belong to this class of oxides. However, the symmetrically equivalent, but oxidation state inverted analogue, the inverse perovskite or antiperovskite (AP) oxide, was much less in the focus of investigation by the scientific community, probably due to its instability in air. The first superconductor among AP oxides, Sr$_{3-x}$SnO, was reported very recently by our group [8]. Potentially, this discovery can be the beginning of the emergence of a new promising family of AP oxide superconductors.

A wide range of oxide compounds exists, crystallizing in the AP and related structures. In general, two formulae for cubic AP oxides are known: $M_3^+O^{2-}A^-$ and $M_3^{2+}A^{4-}O^{2-}$. In both cases O$^{2-}$ is coordinated octahedrally by six $M$, whereas $A$ is coordinated cuboctahedrally by twelve $M$. In the first case, materials are known where $M$ is an alkaline metal or silver and $A$ a halogen or gold [9-18]. In the latter case, materials are known where $M$ is an alkaline earth metal or lanthanoid and $B$ indium or a group 14 element [19-25].

Band structure calculations of AP oxides of the latter case, containing tin or lead as $A$ and alkaline earth metals as $M$, have revealed the presence of six equivalent three dimensional Dirac electrons, located in the very vicinity of the Fermi surface in some of these compounds like Sr$_3$SnO or Ca$_3$PbO [26,27]. The emergence of the Dirac cones is related to a band inversion between the valence band, mainly originating from the $p$ orbitals of tin or lead, and the conduction band, mainly originating from the $d$ orbitals of the alkaline earth metal. These band structures imply nearly closed shell ionic compounds. This is consistent with the observed distances between the alkaline earth and oxygen ions in the crystal structures [8,20]. Thus, the unusual ionic states of Sn$^{4-}$ and Pb$^{4-}$ are realized. Furthermore, it was predicted that a new topological phase, protected by mirror symmetry of the crystal, is present in some members of this family of AP oxides including Sr$_3$SnO [28,29]. Recently, conelike band dispersion in good agreement with the electronic structure calculations [28] were observed by soft x-ray angle-resolved photoemission spectroscopy of Ca$_3$PbO single crystals [30]. This strongly supports that some antiperovskite oxides host 3D Dirac fermions.

The first AP oxide superconductor, Sr$_{3-x}$SnO with a transition temperature $T_c$ of 5 K, was reported very recently by our group [8]. As the parity of the Cooper pairs reflects the orbital texture of the underlying Fermi surface, Sr$_{3-x}$SnO is a candidate for topological superconductivity, due to the mixing of the Sn-5$p$ and Sr-4$d$ bands at the Fermi level. A recent first-priniciples study on the superconducting properties of Sr$_3$SnO predicted it to be a topological superconductor with a $T_c$ of 8.38 K [31]. Early samples of our group, showing superconductivity with volume fractions of up to 32 %, suffered from uncontrolled evaporation of strontium during the synthesis [8]. In fact, superconducting samples contained two cubic phases, which lattice parameters were different by only 0.02 Å. Later we found that the evaporation can be suppressed by synthesizing the samples under argon pressure. A magnetization curve of a sample synthesized under argon pressure is presented in that report. That magnetization curve demonstrates bulk superconductivity with a diamagnetic volume fraction of 62%. However, its synthesis was not discussed and no powder X-ray diffraction (pXRD) data was presented.

In this paper, we present details of the utilization of argon pressure to suppress the strontium evaporation as well as present the synthesis of Sr$_3$SnO from temperatures above its melting point. We present pXRD, energy-dispersive X-ray spectroscopy (EDX), direct current (DC) magnetization and resistivity data for stoichiometric Sr$_3$SnO and deficient Sr$_{3-x}$SnO samples. The analysis of the pXRD data of these improved samples makes it possible to identify the two phases present in the reported superconducting sample, identifying one as SrO and the other one as Sr$_{3-x}$SnO. This is further supported by EDX measurements. The single-phase stoichiometric Sr$_3$SnO samples do not show any superconducting transition down to 0.15 K and exhibit semiconducting behavior in contrast to metallic behavior for the strontium deficient sample in the temperature dependent resistivity. Further, we reveal a strong effect of small amounts of non-magnetic impurities in the strontium starting material on the superconductivity. The magnetic shielding of a strontium deficient "Sr$_{2.5}$SnO" sample exhibits a diamagnetic volume fraction of essentially 100% at 2 K with a superconducting onset of 5.2 K.

## 2. Experimental

*2.1 Sample synthesis*

The samples were synthesized following our previously published procedure (Method A) and with the modified methods B and C.

For all procedures, the strontium (specification below) and the SnO powder (Furuuchi, 99.9%, 169.4 mg, defined as 1.0 molar equivalent) were loaded inside an argon filled glovebox into a crucible. After sealing the crucible in quartz tubes under vacuum (pressure $p <$ 10 Pa) or argon pressure, the quartz tubes were placed at an approximately 45° angle into a box furnace and the appendant temperature program was applied. After the water-quenching, the quartz tubes were opened, stored and prepared for further experiments inside an argon filled glovebox.

*Method A. Synthesis under vacuum.* The synthesis of samples A-1 and A-2 was performed in alumina crucibles (SSA-S, 11×42×1.5 mm) sealed in quartz tubes under vacuum. For the sample A-1 clean cut strontium chunks (Furuuchi, chunks stored in oil, 99.9%, 413.3 mg, 3.75 eq.) and for the sample A-2 clean cut strontium chunks (Furuuchi, chunks stored in oil, 99.9%, 330.6 mg, 3.0 molar equivalent) were used. The sealed quartz tubes were heated to 800°C over 3 h and kept at 800°C for 3 h before water-quenching. In both cases a metallic shining black surface on the inside of the quartz tube wall could be observed, clearly indicating an evaporation of a compound outside of the crucible.

*Method B. Synthesis under argon pressure.* The synthesis of samples B-1, B-2 and B-3 was performed in alumina crucibles (SSA-S, 11×42×1.5 mm) sealed in quartz tubes under 30 kPa of argon pressure at room temperature. For the sample B-1 clean cut strontium chunks (Furuuchi, chunks stored in oil, 99.9%, 331.1 mg, 3.00 eq.), for the sample B-2 clean cut strontium chunks (Furuuchi, chunks stored in oil, 99.9%, 271.6 mg, 2.5 eq.)) and for the sample B-3 dendritic strontium pieces (Sigma Aldrich, dendritic pieces stored under argon, 99.99%, 271.6 mg, 2.5 eq.) were used. The sealed quartz tubes were heated to 825°C over 3 h and kept at 825°C for 3 h before quenching.

*Method C. Synthesis at temperatures above the melting point of $Sr_3SnO$.* The synthesis of the sample C was performed in tantalum crucibles (Japan Metal Service Co. Ltd., 99.95%, 14×35×0.5 mm$^3$) sealed under 23 kPa of argon pressure. Dendritic strontium pieces (Sigma Aldrich, dendritic pieces stored under argon, 99.99%, 341.6 mg, 3.1 eq.) were used. The sealed quartz tubes were heated to 825°C over 40 min and kept at 825°C for 20 min. Subsequently they were heated up to 1200°C in 30 min and kept there for 2.5 h to melt the $Sr_3SnO$ before quenched in water. After the water-quenching, the samples were heated up to 900°C over 1 h and kept there for 48 h to improve the homogeneity and crystallinity before they were quenched in water for a second time.

*2.2 Sample characterization*

*pXRD.* The powder X-ray diffraction measurements for samples B-1 and B-2 were performed at room temperature with a diffractometer (Bruker AXS, D8 Advance) utilizing Cu-K$_α$ radiation (1.54 Å) selected by a nickel-monochromator.

Inside an argon filled glovebox these samples were crushed lightly and the obtained powder placed in the center of the XRD stage, then pressure was applied with a flat glass object to obtain a straight surface and finally the samples were covered with a 12.5-μm-thick polyimide film (DuPont, Kapton) attached to the sample stage by distributing vacuum grease (Dow Corning Toray, high vacuum grease) in the form of a circle around the sample and pressing the polyimide film on top of it.

Samples prepared in this fashion were stable in air for a few hours, before they slowly decomposed into tin and $Sr(OH)_2$. A typical pXRD measurement times was 30 to 220 min.

*Synchrotron pXRD.* Additional pXRD measurements for the sample C were performed at room temperature using synchrotron radiation with a wavelength of $\lambda = 0.42073$ Å on the BL02B2 beam line at SPring-8 (Hyogo, Japan).

The samples were finely ground into powder, and then inserted into a fused-silica capillary tube of inner diameter 0.5 mm under nitrogen atmosphere. During data collection, the tube was continuously rotated to minimize the impact of any preferred orientation, and the scans were obtained over 3 minutes for each sample.

*EDX.* Energy-dispersive X-ray spectra were collected using a commercial detector (AMETEK Inc, model: Element) equipped on a scanning electron microscope (SEM; KEYENCE, model: VE-9800). In a glove box filled with argon, the sample powder was placed on a metallic stage with a conductive double-sided tape made from carbon and aluminum (Nisshin EM Co., Ltd., catalog number: 7321). The samples were exposed to air during the evacuation inside the SEM. Incident electrons with an energy of 15 keV and spot size of 16 were used. To construct the element mappings, we integrated 16 frames for each sample. Every pixel of a frame was obtained by collecting the emitted X-ray for 100 ms.

*2.3 Physical properties*

*MPMS.* A commercial superconducting quantum interference device (SQUID) magnetometer (Quantum Design, MPMS) was used to measure the direct current magnetization $M$.

Samples were covered with grease in an argon environment, then placed inside the straw for measurement. Degaussing was performed prior to measurement to ensure the accuracy of the applied field. Furthermore, the remnant field inside the MPMS was occasionally measured using a reference sample (Pb, 99.9999%), and found to be $\leq 0.1$ mT after degaussing.

*Resistivity.* The resistivity measurement was performed using a four-probe method from 1.8 to 300 K. A sample with a typical size of $2.0 \times 2.0 \times 0.6$ mm$^3$, and 50-μm-diameter gold wires were attached using silver epoxy (EPOXY TECHNOLOGY, H20E) inside an argon filled glovebox. A layer of vacuum grease (Apiezon, N-grease) was applied over the samples inside the glovebox to protect them from contact with air. The samples were cooled during resistivity measurement using a commercial apparatus (Quantum Design, PPMS).

**3. Results and discussion**

*3.1 Synthesis and phase characterization.*

Table 1 gives an overview of the samples presented in this report.

**Table 1.** Summary of the most significant synthesis parameters and the superconducting volume fraction estimated from DC magnetization $M(T)$ data of six different representative batches. No demagnetization correction has been performed on the zero field cooling volume fraction.

| Sample name | Product | Starting composition | Atmosphere at room temperature | $T_{max}$ (°C) | Strontium purity (%) | Superconducting volume fraction at 2 K (%) |
|---|---|---|---|---|---|---|
| A-1 | $Sr_3SnO$ | 3.75Sr + SnO | Vacuum | 800 | 99.9 | <1.5 |
| A-2 | $Sr_{3-x}SnO$ | 3.0Sr + SnO | Vacuum | 800 | 99.9 | 23 |
| B-1 | $Sr_3SnO$ | 3.0Sr + SnO | 0.3 kPa Ar | 825 | 99.9 | 0.0 |
| B-2 | "$Sr_{2.5}SnO$" | 2.5Sr + SnO | 0.3 kPa Ar | 825 | 99.9 | 64 |
| B-3 | "$Sr_{2.5}SnO$" | 2.5Sr + SnO | 0.3 kPa Ar | 825 | 99.99 | 105 |
| C | $Sr_3SnO$ | 3.1Sr + SnO | 0.23 kPa Ar | 1200 | 99.99 | 0.0 |

Samples synthesized under vacuum, as in our previous report (Method A) [8], show a metallic shinning black surface on the inside of the quartz tube. Now we identified this black surface as different $Sr_xSi_y$ phases by pXRD. Therefore, evaporated strontium reduces the $SiO_2$ of the quartz tube to form this dark surface. The amount of strontium evaporating varies from 12% to 25% of the total strontium loading and makes it impossible to control the strontium content in the final product.

We increased the strontium loading to compensate for the evaporated material to synthesize a pure stoichiometric $Sr_3SnO$ phase to compare its superconducting and electrical properties to the ones of $Sr_{3-x}SnO$ of our previous report [8]. In Sample A-1 evaporation and strontium excess compensated each other, as deduced from the final weight of product in the crucible. The pXRD of this sample exhibits only one single cubic phase and no impurities. Therefore, it is considered as stoichiometric $Sr_3SnO$. However, most probably due to a low crystallinity the peaks are very broad, and hence the presence of an additional minor cubic phase with a slightly larger lattice parameter cannot be excluded certainly.

The evaporation can be suppressed through sealing the quartz tubes under an argon atmosphere of 30 kPa at room temperature (Method B), corresponding to 111 kPa at 825°C with an assumption of ideal gas behavior. Due to the successful suppression, it is now possible to control the strontium content in the final product and to investigate the dependence of the superconductivity as well as the emergence of the second cubic phase on the strontium deficiency (Figure 1 (c)). Detailed data on the relation between strontium deficiency and superconductivity will be published elsewhere. Samples synthesized with 3 molar equivalents or more of strontium are single phased and show a high crystallinity (Figure 1 (b)). Decreasing the strontium loading leads to the emergence of a second cubic phase with a 0.02 Å larger lattice parameter (Figure 1 (c)). The difference in the lattice parameters of both phases stays 0.02 Å, independently of the strontium loading. The diffractogram presented in our previous report shows two similar cubic phases. Then, we suspected that this phase splitting is due to different strontium contents in the two different $Sr_{3-x}SnO$ phases. As the new samples of this report synthesized under argon pressure have an improved crystallinity, resulting in narrower peak shapes in the pXRD, the peaks belonging to the two different cubic phases overlap less and can be analyzed more carefully.

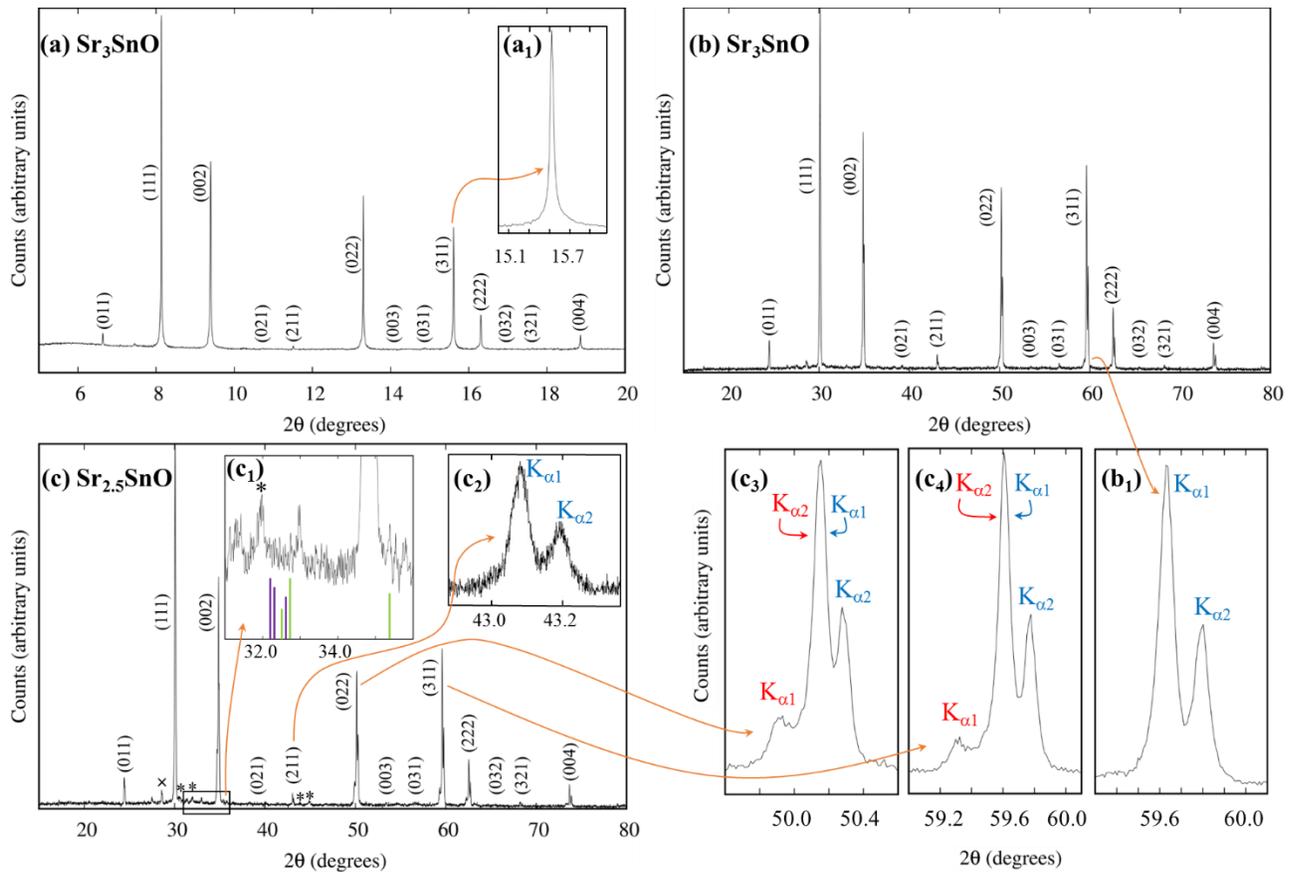

**Figure 1.** Powder X-ray diffraction patterns of $Sr_{3-x}SnO$ at room temperature. The intensities are in linear scales. The diffraction pattern (a) was obtained at the synchrotron-radiation facility SPring-8 with an X-ray wavelength of $\lambda = 0.42073$ Å. The diffraction patterns (b) and (c) were obtained with a laboratory-based diffractometer utilizing Cu-K$_\alpha$ ($\lambda = 1.54$ Å) radiation. (a) and (b) are for stoichiometric and non-superconductive $Sr_3SnO$ of Sample C and B-1, respectively. (c) is for the superconductive "$Sr_{2.5}SnO$" Sample B-2. The insets are magnifications of certain peaks or areas. The green and purple bars in inset ($c_1$) indicate the expected position and intensity (PDF-2 database: PDF010734901 for $SrSn_3$ and PDF010723942 for $SrSn_4$) of the three most pronounced peaks of the only two known superconducting Sr-Sn-alloys $SrSn_3$ and $SrSn_4$, respectively. For the diffractograms of the stoichiometric batches (a) and (b) only the primitive cubic $Sr_3SnO$ phase can be observed. For (b) and (c) a splitting of the peaks due to $K_{\alpha 1}$ and $K_{\alpha 2}$ radiation is present. The additional splitting in the peaks of (c) is due to an additional face centered cubic SrO phase. The peaks originating from the SrO phase ($a = 5.160$ Å) are indicated by $K_{\alpha 1\text{ or }2}$ in red in the insets ($c_3$) and ($c_4$). The ones originating of the "$Sr_{2.5}SnO$" phase ($a = 5.137$ Å) are indicated by $K_{\alpha 1\text{ or }2}$ in blue in the insets ($c_2$), ($c_3$), ($c_4$) and ($b_1$). In inset (c) peaks of a SnO and tin impurity (PDF-2 database: PDF000130111 for SnO and PDF030650296 for tin) are marked by an x and asterisks, respectively.

A diffractogram of such a new Sample B-2 is shown in Figure 1 (c). The lattice parameter of the main cubic phase is 5.137 Å and the one of the minor phase 5.160 Å. Recently, a lattice parameter of 5.139 Å for $Sr_3SnO$ single crystals was reported [25]. Hence, agreeing well with our main phase. The lattice parameter of the minor phase agrees with 5.160 Å, reported for SrO [32]. Additionally, the peak intensity ratios of the (1 1 1), (2 0 0), (2 2 0) and (3 1 1) peaks are expected to be different in $Sr_3SnO$ and SrO. The leftmost peak of the three peaks present for each of these four diffraction planes belongs exclusively to the minor phase, see the peaks marked with red $K_{\alpha 1}$ in Figure 1 ($c_3$) and ($c_4$) as an example for the (2 2 0) and (3 1 1) diffraction planes. The intensity ratios of these leftmost peaks fit to the one expected for SrO and the one of the rightmost peaks, originating exclusively from the major phase, to the one of $Sr_3SnO$. To gain further support for the presence of SrO, one of the low intensity peaks (2 1 1), expected to be only present in the primitive cubic $Sr_3SnO$ and not in face centered cubic SrO, was measured exclusively overnight; see magnified peak in Figure 1 ($c_2$). No splitting in two phases is present in this peak, indicating that no second primitive cubic, but a face centered cubic phase is present. Furthermore, the problem of alkaline earth oxide impurities is reported in literature known for the synthesis of bulk [23,25] and thin film [33] antiperovskite oxides. Therefore, we conclude that the face centered cubic phase is SrO.

One SnO (superconducting only under pressure) [34] and four very small tin ($T_c = 3.7$ K) [35] impurity peaks in Figure 1 (c) are marked with an x and asterisk, respectively (PDF-2 database: PDF000130111 for SnO and PDF030650296 for tin). However, due to the small intensity an unambiguous identification by pXRD is not possible. The EDX results presented in Figure 2 (a) show a small area containing only tin. Additionally, the DC magnetization $M(T)$ curves occasionally show a very small anomaly around 3.7 K. Hence, we conclude a small tin impurity is very likely.

Due to the SrO and tin impurity the precise stoichiometry of the final AP oxide phase is challenging to determine. The strontium deficiency might be slightly bigger than expected from the initial strontium loading. Furthermore, an oxygen deficiency is possible. However, considering that $Sr_3Sn$ without oxygen could not be synthesized despite attempts during work on Sr-Sn phase diagrams [19,20], a significant oxygen deficiency is unlikely. It is possible that small amounts of oxygen are provided through a reaction with the alumina crucible or from another unknown source. Due to these uncertainties in strontium deficient sample, their molecular formula is set into quotation marks in this report.

Two Sr-Sn-alloys are known to be superconducting, $SrSn_3$ ($T_c$ = 5.4 K) [36] and $SrSn_4$ ($T_c$ = 4.8 K) [37]. The green and purple bars in inset ($c_1$) of Figure 1 indicate the expected position and intensity (PDF-2 database: PDF010734901 for $SrSn_3$ and PDF010723942 for $SrSn_4$) of the three most pronounced peaks of $SrSn_3$ and $SrSn_4$, respectively. The peak positions don't match to those of the experimental pXRD data. Hence, the presence of crystalline $SrSn_3$ and $SrSn_4$ can be excluded.

Scanning electron microscopy (SEM) and energy-dispersive X-ray (EDX) spectroscopy was performed on "$Sr_{2.5}SnO$" of Sample B-2, see Figure 2 (a). Unfortunately, during some parts of the procedure we didn't succeed in protecting the samples from air. Therefore, we couldn't maintain flat surfaces and reliable quantitative results for the element ratios. However, the EDX images clearly show that the distributions of oxygen (red), tin (blue) and strontium (green) are very similar in a major part of the samples surface, supporting the dominance of the $Sr_{3-x}SnO$ phase and excluding a significant amorphous impurity with a different composition. The orange circles mark an area, containing only tin and the white circles an area containing strontium and oxygen. These EDX results are in accordance to our pXRD data, see Figure 1 (b), which shows $Sr_{3-x}SnO$ as dominant phase with SrO oxide and tin as impurities. In general, much more areas containing only strontium and oxygen could be identified by EDX than areas containing only tin, as also expected from the pXRD data.

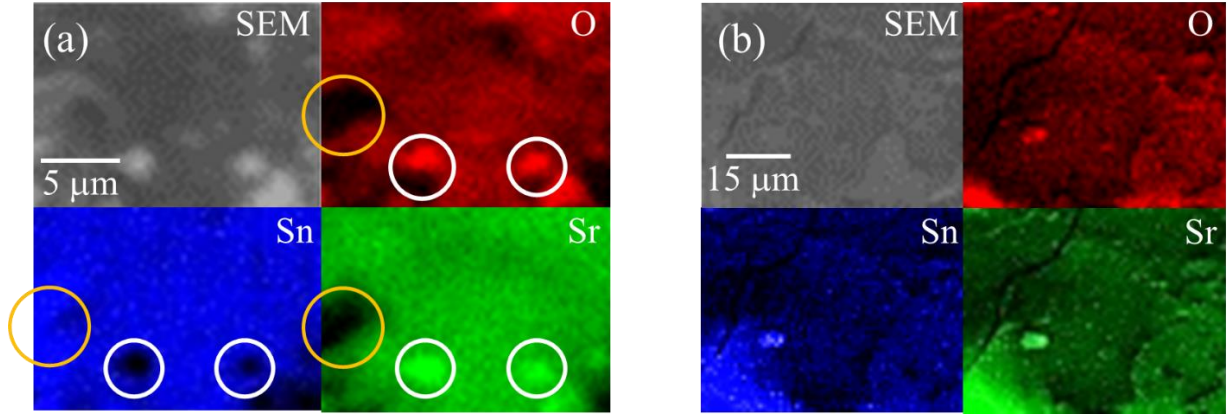

**Figure 2.** SEM and EDX images of "$Sr_{2.5}SnO$" of Sample B-2 (a) and "$Sr_3SnO$" of Sample C (b). The air sensitive samples were exposed to air while vacuum was applied to the SEM/EDX chamber. Therefore, no flat surfaces and reliable quantitative analysis could be obtained. The grey image at the top left side of (a) and (b) is the SEM image and the other three are EDX images of the same area. The oxygen distribution is red, the tin one blue and the strontium one green. All three elements are distributed in a similar way in (a) and (b), indicating a strong dominance of the $Sr_{3-x}SnO$ phase. In (a) are some areas with different element ratios, indicating impurities. The white circles in (a) indicate a region containing only strontium and oxygen. The orange circles an area containing only tin.

A brief screening of reaction temperatures in the range of 800°C to 1000°C and reaction times from 1 h to 48 h was performed. No significant differences could be observed in the pXRD and DC magnetization $M(T)$ data. We choose a maximum furnace temperature of 825°C instead of 800°C, as it guaranties the complete melting of the strontium ($T_m$ = 777°C) in our setup and a reaction time of 3 h seems most suitable.

The stoichiometric sample A-1 still shows a superconducting volume fraction of less than 1.5% at 2 K, indicating inhomogeneity of the strontium distribution with some deficient domains. To obtain more homogeneous materials with an improved crystallinity, $Sr_3SnO$ was synthesized by heating the stoichiometric starting materials above $T_m(Sr_3SnO)$ = 1080°C and with an argon pressure of 22 kPa at room temperature. However, the molten $Sr_3SnO$ reacts with the alumina crucible, forming $Sr_x(Al_yO_z)$. Thus, tantalum crucibles were used instead. Initially, slow cooling through the melting point was tried to grow single crystals. However, crystals with a sufficient size were not obtained and noticeable amounts of SrSn ($T_m$ = 1140°C) as well as SrO were present in all these samples. Therefore, a different method was used in which samples were water-quenched from 1200°C and then sintered at 900°C for 48 h to improve their crystallinity. Samples synthesized in this way with stoichiometric or slightly over stoichiometric strontium loadings do not show any superconductivity down to 0.15 K in alternating current (AC) susceptibility measurements, indicating a high homogeneity in the strontium distribution. A high energy pXRD of such a sample C is shown in Figure 1 (a). It is free of any impurities and the peak shape indicates high crystallinity. Figure 2 (b) shows EDX images of the oxygen (red), tin (blue) and strontium (green) distribution. Unfortunately, as with sample B we didn't succeed in protecting the samples from air during the whole procedure. Therefore, we couldn't obtain flat surfaces and reliable quantitative results for the element ratios. All three elements are distributed in the same way supporting the presence of a clean $Sr_3SnO$ phase, as indicated by the pXRD data of Figure 1 (a).

As quenching $Sr_3SnO$ from above its melting point was successful for synthesizing stoichiometric samples with a homogenous strontium distribution, we attempted the same method for strontium deficient samples. However, we found that at temperatures above the melting point deficient $Sr_{3-x}SnO$ decomposes into $Sr_xSn_y$, SrO and stoichiometric $Sr_3SnO$. Hence, this procedure is not suitable for the synthesis of superconducting strontium deficient samples. Furthermore, this observation indicates that deficient $Sr_{3-x}SnO$ is metastable even under an inert atmosphere, making it challenging to synthesize.

Another crucial aspect for the synthesis of superconducting $Sr_{3-x}SnO$ is a high purity of the strontium raw material, as discussed in the following section.

*3.2 Superconductivity and normal state properties.*

Figure 3 shows DC magnetisation data $M(T)$ of stoichiometric $Sr_3SnO$ (Sample B-1), of "$Sr_{2.5}SnO$" with 99.9% purity of the strontium substrate (Sample B-2) and "$Sr_{2.5}SnO$" with 99.99% purity of the strontium substrate (sample B-3). Despite the strontium purity the synthesis method of B-2 and B-3 is identical and the pXRD data shows no significant differences. The curve for the zero-field-cooling (ZFC) process of Sample B-2 shows a magnetic flux expulsion corresponding to an apparent superconducting volume fraction of 64% without demagnetization correction at 2 K with an onset of 4.8 K. The ZFC curve of Sample B-3 reveals a superconducting volume fraction of 105% with an onset of 5.2 K. Both superconducting volume fractions clearly indicate bulk superconductivity of the main "$Sr_{2.5}SnO$" phase. A weak effect on the $T_c$ and a significant dependency of the superconducting volume fraction on the strontium purity can be deduced. Remarkably, for "$Sr_{2.5}SnO$" samples synthesized like B-2 and B-3, but with strontium of 99% purity[1], no superconducting transition in the DC magnetization $M(T)$ down to 2 K can be observed. Such a high sensitivity towards nonmagnetic impurities is known for some unconventional superconductors such as $Sr_2RuO_4$ [38]. The field-cooling (FC) curves in Figure 3 of the samples B-2 and B-3 are much less pronounced, as expected for a type-II superconductor due to flux pinning. In the ZFC curve of $Sr_3SnO$ (Sample B-1) no sizeable diamagnetic transition is present down to 2 K. Furthermore, AC susceptibility measurements couldn't reveal a superconducting transition down to 0.15 K. These results clearly indicate that only strontium deficient $Sr_{3-x}SnO$ is superconducting.

---

[1] Rare Metallic, impurities according to the manufacture: Ca < 0.13%, Ba < 0.16%, Mg: 0.005%, Si: 0.0021%, Fe: 0.0008%, Na: 0.0026%.

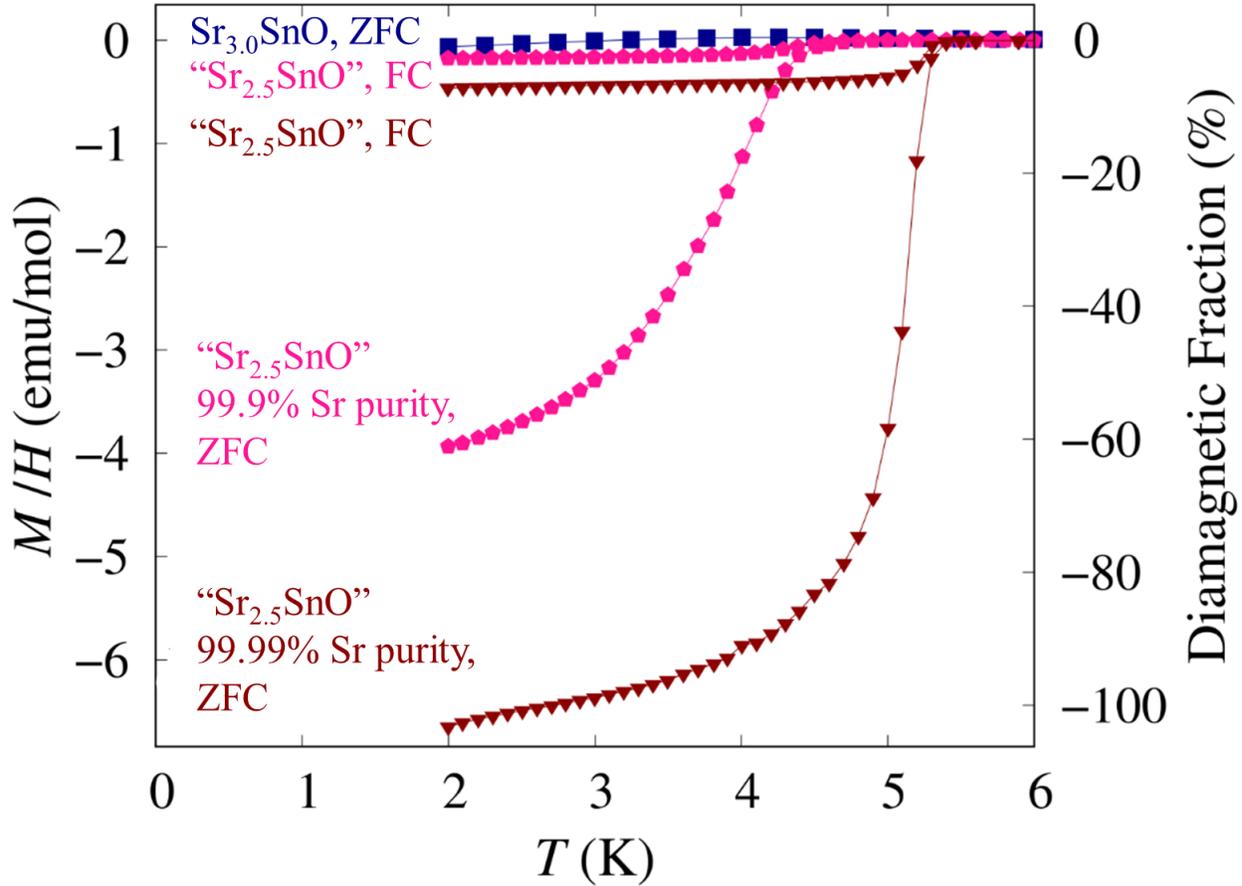

**Figure 3.** Magnetization under zero field cooling (ZFC) with an applied field of 2 mT of $Sr_3SnO$ of Sample B-1 (blue square), and under ZFC as well as field cooling (FC) of "$Sr_{2.5}SnO$" of Sample B-2 (pink pentagon) and Sample B-3 (brown triangles). The vertical scale on the right indicates the estimated diamagnetic volume fraction without demagnetization correction. Stoichiometric $Sr_3SnO$ is not superconducting. The superconducting volume fraction of "$Sr_{2.5}SnO$" increases significantly with an increased purity of the strontium raw material.

The blue curve in Figure 4 shows the dependence of the electrical resistivity $\rho$ on the temperature of polycrystalline chunks of $Sr_3SnO$ (Sample A-1). It indicates weak semiconducting behavior with the value at 10 K to be 0.154 $\Omega$cm. This is consistent with the first-principles band calculations of our recent report, implying an isolated gapped Dirac cone in the very vicinity of the Fermi energy [8]. The presence of another hole pocket around the R point, as indicated by a first-principles band calculation by Hsieh *et al.* [28], stands in contrast to this result. The absence of such a hole pocket around the R point further supports our claim that the formation of the Cooper pairs appears on the Fermi surface originated from the Dirac points, and therefore could results in unconventional superconducting properties [8]. The temperature dependence of $\rho$ is metallic for $Sr_{3-x}SnO$ (Sample A-2, Figure 4 red curve) with the value at 10 K to be 65 $\mu\Omega$cm. This is expected, as the Fermi energy should be shifted downwards, due to the heavy hole doping caused by the strontium deficiency. For this sample $\rho$ drops sharply with an onset of 4.8 K and reaches zero at 4.5 K, and thus further supports the claim of bulk superconductivity for strontium deficient $Sr_{3-x}SnO$. For $Ca_3SnO$ and $Ca_3PbO$, predicted to have a similar band structure with small band gaps [28], metallic temperature dependence of $\rho$ was also observed for samples synthesized under vacuum [39]. In the $Sr_3SnO$ sample A-1, the resistivity seems to saturate at ~25 K followed by a slight drop down to ~5 K and a sharp drop at 3.8 K. This drop is most likely due to a small inclusion of metallic tin ($T_c$ = 3.7 K) in our sample. The scattering is expected to increase in the hole-doped $Sr_{3-x}SnO$ sample, but the increases in number of carriers with doping is more significant, which is demonstrated by the 1000 times decrease in $\rho$ of the metallic $Sr_{3-x}SnO$ sample compared with the semiconducting $Sr_3SnO$ sample. These results show the significant effect of doping in this material, and may arise from the Dirac dispersion observed in the band calculations.

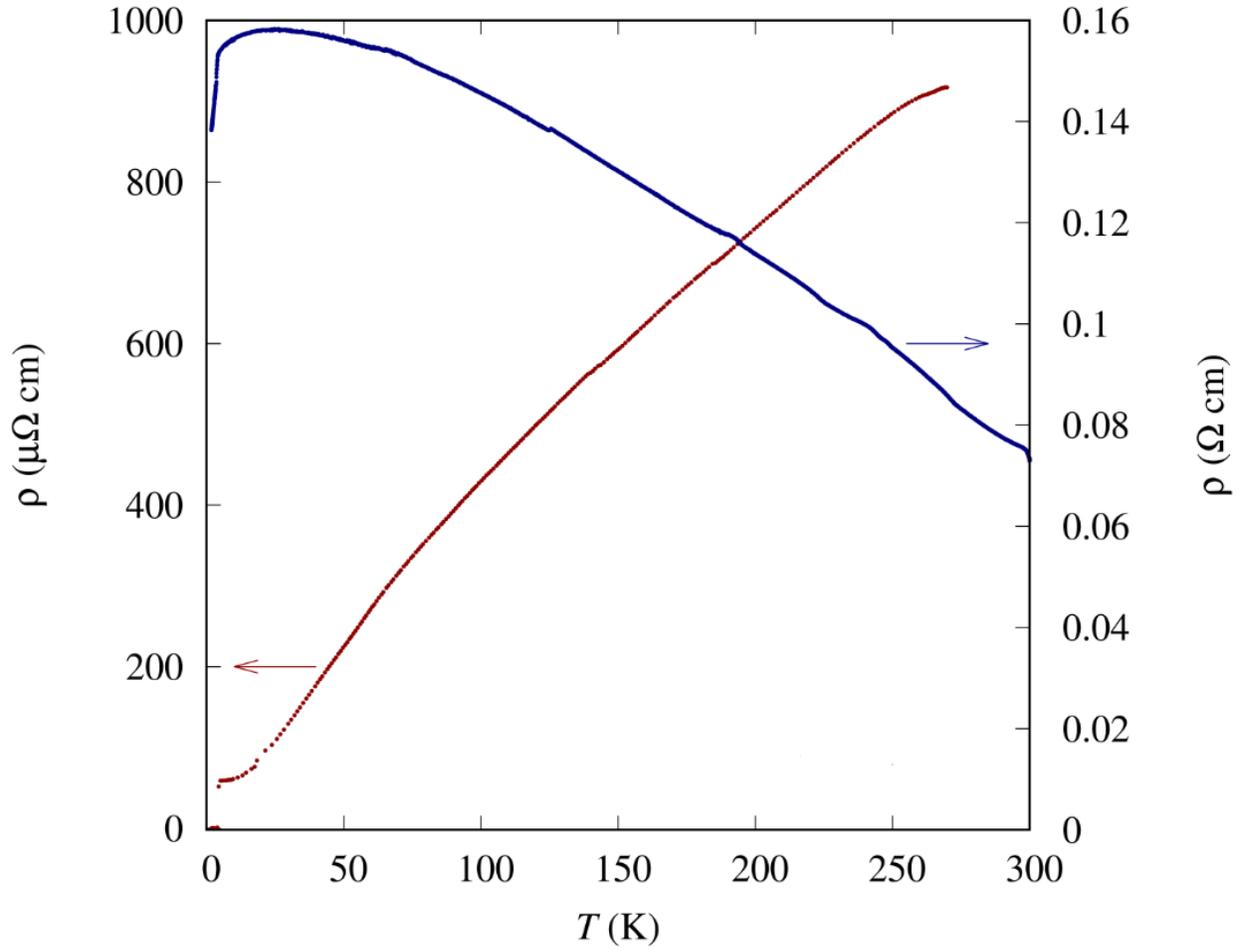

**Figure 4.** Electrical resistivity $\rho$ as a function of temperature of polycrystalline chunks of $Sr_3SnO$ of Sample A-1 in blue (arrow to the right) and of $Sr_{3-x}SnO$ of Sample A-2 in red (arrow to the left). For stoichiometric $Sr_3SnO$, $\rho(T)$ is semiconducting with the value at 10 K to be 0.154 $\Omega$cm. The drop at 3.8 K is most likely due to a metallic tin impurity ($T_c$ = 3.7 K). For hole doped $Sr_{3-x}SnO$, $\rho(T)$ is metallic with the value at 10 K to be 65 $\mu\Omega$cm. This behavior is in accordance with our recent band structure calculations [8].

## 4. Conclusion

The synthesis Method B, presented in this report, at 825°C for 3 hwith an argon pressure of at least 30 kPa at room temperature, a molar Sr:SnO ratio of 2.5 and a strontium raw material purity of 99.99% succeeds in producing bulk polycrystalline $Sr_{3-x}SnO$ with a superconducting volume fraction of around 100%. Furthermore, the magnitude of the strontium deficiency can be controlled and we demonstrate the absence of superconductivity in stoichiometric $Sr_3SnO$. Additionally, we performed a more detailed characterization of the superconducting "$Sr_{2.5}SnO$" material. With this synthesis method established and the improved understanding of the material the nature of the superconductivity and the dependence of it on the strontium deficiency in $Sr_{3-x}SnO$ can be investigated more detailed in the future.

**Acknowledgments**

We acknowledge Cedric Tassel, Fumitaka Takeiri, Hiroshi Takatsu and Hiroshi Kageyama for their contribution to the synchrotron pXRD measurements. We also thank S. Kasahara, Y. Matsuda and M. Maesato for their support. We thank Nicola Pinna for his discussions.

This work was supported by the JSPS KAKENHI Nos. JP15H05851, JP15H05852, and JP15K21717 (Topological Materials Science), by the JSPS Core-to-Core program, as well as by Izumi Science and Technology Foundation (Grant No. H28-J-146).

NH is supported by the Kyoto inter-university exchange program.

AI is supported by Japan Society for the Promotion of Science as JSPS Research Fellow (Grant No. JP17J07577).